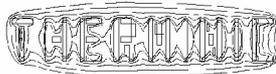

# Flexible Profile Approach to the Steady Conjugate Heat Transfer Problem

M.-N. Sabry
Université Française d'Égypte
Shourouk
Cairo, 11837 Egypt

*Abstract*- The flexible profile approach proposed earlier to create CTM (compact or reduced order thermal models) is extended to cover the area of conjugate heat transfer. The flexible profile approach is a methodology that allows building a highly boundary conditions independent CTM, with any desired degree of accuracy, that may adequately replace detailed 3D models for the whole spectrum of applications in which the modeled object may be used. The extension to conjugate problems radically solves the problem of interfacing two different domains. Each domain, fluid or solid, can be "compacted" *independently* creating two CTM that can be joined together to produce reliable results for any arbitrary set of external boundary conditions.

## I. Problems Associated with the Heat Transfer Coefficient

Conjugate heat transfer is a classical problem in heat transfer in which coupling between two heat transfer modes (convection and conduction) is observed. This phenomenon was raised to the level of a "problem" because of the non adapted use of the concept of heat transfer coefficient $h$ and associated correlations to this case.

In fact, at the boundaries of a fluid undergoing forced convection, heat transfer boundary conditions are in reality never as ideal as those used to obtain correlations, i.e. uniform temperature or uniform heat flux. The concept of heat transfer coefficient $h$ represents a compact thermal model, although of the lowest possible order. It reduces (or compacts) the complicated 3D problem to a simplified form containing only one parameter (or one degree of freedom): a single resistance. Advantages of simplicity are evident. Disadvantages are rarely acknowledged. The most important one being that this simple model can never be Boundary Conditions Independent (BCI). Indeed, we usually say the "uniform heat flux $h$," or the "uniform temperature $h$" etc. One should never use a non BCI model in situations other than those for which it was extracted, which is unfortunately rarely respected in engineering practice, simply because no other solution is offered. It would of course be impractical to produce dozens of correlations for $h$ for the same geometry but with different forms of boundary conditions. The case of conjugate heat transfer is a flagrant manifestation of the problem: Even if we had a whole set of correlations for $h$ for all possible forms of boundary conditions (uniform, non-uniform, imposed temperature, imposed flux …), we would have problems selecting the appropriate element in this set because the actual temperature and heat flux profiles are unknown. The only plausible solution to that problem is to replace the heat transfer coefficient $h$ by another BCI representation of the heat transfer problem. This constitutes the main objective of this work.

Compact thermal model (CTM) that model conduction problems associated with heat transfer in electronic industry is a subject that has received considerable attention [1-4] and is thus sufficiently mature. It has long been recognized in this area that the "equivalent" thermal resistance of a domain undergoing heat transfer by conduction is a strong function of temperature and/or heat flux profiles at its boundaries [5]. The BCI concept has first been introduced in this field [6]. Of course no CTM can be 100% BCI, but care in creating it would let it be as close to BCI as possible.

Concerning dependence of the thermal resistance on imposed profiles, the situation for convection is even worse: the genesis and development of the thermal boundary layer is a strong function of these profiles. Conjugate heat transfer problems are the worst of all simply because profiles are far from being uniform as well as being unknown. Another approach was also proposed by solving (or measuring) the coupled problem in order to derive correlations taking into consideration properties of both media. This leads to an intractable problem because of the huge number of possible combinations of convective and conductive domains.

All this is also linked with the well known fact that the concept of heat transfer coefficient does not adequately model multiple heat source problems [7]. If multiple heat sources were installed on domain boundaries, with the ability to operate independently, any form of boundary conditions profiles can be created. They cannot be all modeled by one and the same correlation. The use of a "local" heat transfer coefficient does not help, simply because the development of the thermal boundary layer is not a local phenomenon.

## II. Inability of Resistive Networks to a Convective Domain

In an earlier work [8], it was argued that resistive networks are by nature fully symmetric, in the following sense: A source acting on node $i$ will produce an effect on node $j$ that is exactly the same as the effect that would have been produced at node $i$ if the source was placed at node $j$. This is perfectly valid for conduction heat transfer, because it is governed by a differential equation (3.1a) containing a symmetrical operator. Convection is precisely the case where this statement is not true! A source placed up-stream will produce an effect on down stream side that is orders of magnitude higher than the effect that would have been produced up-stream by the same source placed down stream. This well known fact stems from the non-symmetrical nature of the differential operator that appears in the energy equation describing convection (4.1b). Above argument is a killing one for any attempt to model a domain undergoing convection by a resistive network. In order to put it in a visual form, imagine a micro-channel with fluid entering at a low temperature $T_{in}$, leaving at a higher temperature $T_{out}$ and exchanging heat with a wall that is maintained at a uniform temperature $T_{wall}$, which is the highest temperature. There are obviously 3 nodes in this model *in*, *out* and *wall*. A fourth one may be also defined, with which we can associate the average fluid bulk temperature $T_{bulk}$. In addition to the temperature associated with each node, we have to associate an "energy flux": $q_{in}$, $q_{out}$ and $q_{wall}$. The first two contain mainly (but not exclusively) an enthalpy flux, the last one only contain heat crossing the wall. The reason why we consider energy fluxes, and not only heat fluxes, is because heat entering walls in this problem leaves in the form of enthalpy. The role of the CTM is precisely to model the relation between all these energies and corresponding temperatures as a black box. We need not defining an independent energy flux associated with the bulk node, unless heat was generated in the bulk, which is a case we preclude for simplicity. How can we build the CTM? In Figure 1, all nodes are shown, together with real directions of energy fluxes. A single resistance that is simply the inverse of *h* is also shown. Can we connect *in/out* nodes with other nodes, using resistors? Certainly not, because such resistors would imply heat transfer from *out* to *in* through the *bulk*! Any attempt to "solve" this issue using only resistors would simply fail because resistors are not valid building blocks for convective problems. In case the convection problem was linear, the **_only possible_** way to model *T/q* relations is a non-symmetric matrix (contrary to the conductive case which results in a symmetric matrix).

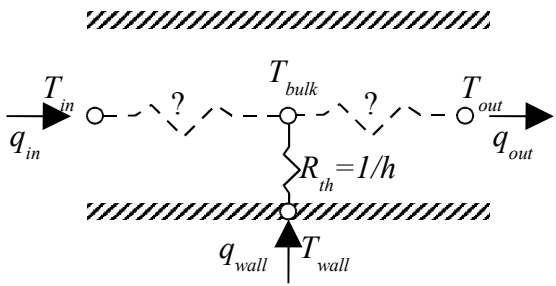

Figure 1. Inability of a resistive network to model a convective domain

### III. THE FLEXIBLE PROFILE TECHNOLOGY

The flexible profile approach has been recently introduced [9] to model heat transfer by conduction under the effect of any type, form or value of boundary conditions. This is a fundamentally new approach that performs much better than the classical "equivalent thermal resistance(s)" one. The main philosophy is simple:

A. Construct a set of basic profiles that may serve as a basis to generate _any_ profile. It has been proven that this set is composed of relatively few elements, of the order of 1 to 5 per direction for each "node" (i.e. heat exchange port).

B. Obtain analytically the response of the system to each element of the set, in the form of a matrix of influence coefficients, which is a generalization of the thermal resistance network. This is the new Compact Thermal Model (CTM).

C. Use the above obtained CTM to predict system response for any given profile, by decomposing this profile over the basis constructed in step A. The more elements in the basis the better would be the representation of the imposed boundary conditions. This is the only approximation involved. The error is perfectly controllable by selecting any adequate number of elements in the basis constructed in step A.

The detailed model taking into consideration only classical conduction effects (i.e. no quantum effects) at steady state is governed by:

$$\nabla \cdot (\lambda \nabla T(\mathbf{r})) = -q_v(\mathbf{r}) \quad \mathbf{r} \in \Omega \quad (3.1a)$$

$$T|_{\partial\Omega_D} = T_s ; \quad \lambda \mathbf{n} \cdot \nabla T|_{\partial\Omega_N} = q_s ; \quad (3.1b)$$

where $T$ is the temperature field, $q_v$ is a volumetric heat generation term, $\lambda$ is the thermal conductivity, $\mathbf{r}$ is the coordinate, $q_s$ is the heat flux density entering the surface. The latter can be expressed as any linear or non-linear function of $T$, allowing thus Robin type boundary conditions:

$$q_s = h(T_\infty - T) \quad (3.1c)$$

(where $h$ is the heat transfer coefficient by convection) as well as radiation and non-linear convection. In the sequel, $\lambda$ will be considered constant in order to have a linear PDE (3.1a). This is not a limitation, since the case where $\lambda$ is function of $T$ can be easily transformed to a linear equation using Khirchhoff transformation. Note that boundary conditions may still be a nonlinear function of $T$ through $h$ in (3.1c). The application of the flexible profile approach will be very briefly presented. First, let us create an associated modified Green's function $G$ satisfying:

$$\nabla^2 G(\mathbf{r},\mathbf{r}') = -\delta(\mathbf{r}-\mathbf{r}') \quad \mathbf{r}, \mathbf{r}' \in \Omega \quad (3.2a)$$

$$\mathbf{n} \cdot \nabla G(\mathbf{r},\mathbf{r}') = \begin{cases} -1/\int_{\mathbf{r} \in \partial\Omega_1} d\mathbf{r} & \mathbf{r} \in \partial\Omega_1 \\ 0 & \mathbf{r} \notin \partial\Omega_1 \end{cases} \quad (3.2b)$$

$$\int_{\mathbf{r} \in \partial\Omega_1} G d\mathbf{r} = 0 \quad (3.2c)$$

where $\partial\Omega_1$ is the outside surface sub-domain corresponding to an arbitrarily chosen "reference" node. Using the above Green's function, as well as the Green's theorem, equation (3.2a) can be transformed into:

$$T(\mathbf{r}) - T_{av1} = \sum_{j=1}^{N} \int_{\mathbf{r'}\in\Omega_j} G(\mathbf{r},\mathbf{r'})q(\mathbf{r'})d\mathbf{r'} \qquad (3.3)$$

where $T_{av1}$ is the average temperature over node 1 and $\Omega_j$ is the domain of "node" $j$. Using any convenient orthonormal set $\phi_i^u(\mathbf{r})$ over each node $i$, we can express $T$ and $q$ profiles as:

$$T(\mathbf{r})|_{\mathbf{r}\in\Omega_i} = \sum_{u=0}^{\infty} T_i^u \phi_i^u(\mathbf{r}) \qquad (3.4a)$$

$$q(\mathbf{r})|_{\mathbf{r}\in\Omega_i} = \sum_{u=0}^{\infty} q_i^u \phi_i^u(\mathbf{r}) \qquad (3.4b)$$

Hence, by substituting in (3.3) we get after multiplying both sides by $\phi_i^u(\mathbf{r})$, integrating using the orthonormal property of $\phi_i^u(\mathbf{r})$ and truncating the series after $U$ terms:

$$T_i^u = \sum_{j=1}^{N}\sum_{v=0}^{U} R_{ij}^{uv} q_j^v \quad i\in[1,N], u\in[0,U] \qquad (3.5a)$$

$$R_{ij}^{uv} = \int_{\mathbf{r}\in\partial\Omega_i}\int_{\mathbf{r'}\in\partial\Omega_j} \phi_i^u(\mathbf{r})G(\mathbf{r},\mathbf{r'})\phi_j^v(\mathbf{r'})d\mathbf{r'}d\mathbf{r} \qquad (3.5b)$$

This is the Flexible Profile Compact Model involving as state variables at each node, not only one single value, but rather the coefficients of the expansion of $T$ and $q$ profiles over a given complete set. Taking a sufficient order of precision over each node, typically 0 to 4 per dimension, one can approximate reasonably well any $T$ or $q$ profile with one and the same model. It is worth noting that classical approaches are explicitly or implicitly equivalent to the restricted version of the proposed approach with only one term in the series (3.4). The use of the series (3.4) as well as the general equation (3.3) makes the flexible profile approach a general framework for all possible CTM in conduction

## IV. EXTENSION TO THE CONJUGATE PROBLEM

The concept depicted above will now be generalized to the case of forced convection, before extending it to the conjugate problem, both being the main contribution in this work. We limit analysis to forced convection with a known velocity field $\mathbf{v}$ to maintain linearity. It is to be noted that non-linearities in external boundary conditions will still be easily handled. Governing differential equation is now:

$$\nabla\cdot(\lambda\nabla T(\mathbf{r})) - \rho c_p \mathbf{v}\cdot\nabla T(\mathbf{r}) = -q_v(\mathbf{r}) \quad \mathbf{r}\in\Omega \qquad (4.1a)$$

Boundary conditions will be expressed in a form that reveals energy fluxes as follows:

$$T|_{\partial\Omega_D} = T_s ; \quad (\lambda\mathbf{n}\cdot\nabla T - \rho c_p T\mathbf{v}\cdot\mathbf{n})|_{\partial\Omega_N} = q_s ; \qquad (4.1b)$$

As before, the energy flux density $q_s$ can take any form, including a linear (for convection: eq. 3.1c) or nonlinear (for radiation) relation with temperature. Physical properties are constant. The associated modified Green function satisfies:

$$\lambda\nabla^2 G(\mathbf{r},\mathbf{r'}) + \rho c_p \mathbf{v}\cdot\nabla G = -\delta(\mathbf{r}-\mathbf{r'}) \quad \mathbf{r},\mathbf{r'}\in\Omega \qquad (4.2a)$$

$$\mathbf{n}\cdot\nabla G(\mathbf{r},\mathbf{r'}) = \begin{cases} -B/\int_{\mathbf{r}\in\partial\Omega_1} d\mathbf{r} & \mathbf{r}\in\partial\Omega_1 \\ 0 & \mathbf{r}\notin\partial\Omega_1 \end{cases} \qquad (4.2b)$$

where $\partial\Omega_1$ is the outside surface sub-domain corresponding to an arbitrarily chosen "reference" node and $B$:

$$B = \left(1 + \int_{\mathbf{r}\in\partial\Omega}\rho c_p \mathbf{v}\cdot\mathbf{n}G d\mathbf{r}\right) \qquad (4.2c)$$

In case the integral in the RHS of (4.2c) vanishes (e.g. conduction), the following condition could be added to remove the arbitrary additive constant when all boundary conditions involve derivatives:

$$\int_{\mathbf{r}\in\partial\Omega_1} G d\mathbf{r} = 0 \qquad (4.2d)$$

Using the above Green's function, as well as the Green's theorem, equation (4.2a) can be transformed into:

$$T(\mathbf{r}) - T_{av1m} = \sum_{j=1}^{N}\int_{\mathbf{r'}\in\Omega_j} G(\mathbf{r},\mathbf{r'})q(\mathbf{r'})d\mathbf{r'} \qquad (4.3)$$

where $T_{av1m}$, defined as:

$$T_{av1m} = \int_{\mathbf{r}\in\partial\Omega_1} BT d\mathbf{r} \Big/ \int_{\mathbf{r}\in\partial\Omega_1} d\mathbf{r}$$

is an "average" temperature over node 1 with the weight $B$. As for $q$, it can be either $q_v$ if the node $j$ was a volume source or $(\lambda\mathbf{n}.\nabla T - \rho c_p T \mathbf{v}.\mathbf{n})$ if node $j$ was a surface source.

Equation (4.3), derived in this work, plays for convective domains, the same role as that played by (3.3) for conductive domains. It is a general relation between energy fluxes and temperatures at system boundaries that was analytically obtained without any simplifying assumption, and without making use at any level of boundary conditions (4.1 b, c). This fully BCI model is the general framework for any CTM to be built for convective domains. In analogy with (3.3) for conduction, any CTM constructed for convective domains can be viewed as an approximation to (4.3) in order to reduce its dimensionality. The reduction will naturally entail some loss of its BCI properties.

Under the light of what has been advanced, a rational approximation which keeps any desired level of BCI properties of the resulting CTM would be to consider its truncated development over a complete orthonormal set (e.g. 3.4). This will give rise to:

$$T_i^u = \sum_{j=1}^{N}\sum_{v=0}^{U} R_{ij}^{uv} q_j^v \quad i\in[1,N], u\in[0,U] \qquad (4.4a)$$

$$R_{ij}^{uv} = \int_{\mathbf{r}\in\partial\Omega_i}\int_{\mathbf{r'}\in\partial\Omega_j} \phi_i^u(\mathbf{r})G(\mathbf{r},\mathbf{r'})\phi_j^v(\mathbf{r'})d\mathbf{r'}d\mathbf{r} \qquad (4.4b)$$

The above equation is identical to (3.5), except that $T$, $q$ and $G$ here are generalized entities that may reduce to their conductive counterparts if $\mathbf{v}$ vanishes everywhere ($\Rightarrow B=1$):

$T$     temperature with reference $T_{av1m}$ (instead of $T_{av1}$)
$q$     Energy flux density (instead of heat flux density)
$G$     Green's function satisfying (4.2) instead of (3.2)

The formulation obtained in this work goes beyond earlier suggestion [8] due to the high analogy between CTM formats obtained for convective and conductive domains, which is quite adapted to the conjugate heat transfer problem.

The matrix *R* appearing in (4.2) would adequately replace the single heat transfer coefficient *h*. It is interesting to note that the *R* matrix contains *h* for the uniform heat flux case. The inverse of *R* contains *h* for the uniform temperature case. Hence, we can easily visualize the fact that *R* gives the full picture that a single *h* can not capture. The price paid by this complication, i.e. replacing a single number by a matrix, is largely compensated by the ability to model any type of boundary conditions. In fact, the CTM constructed this way, describes precisely the intrinsic behavior of each domain (convective or conductive) subject to any external effect.

The Green's function *G* for convective domains may not be easy to find analytically for complicated geometries. These geometries can be partitioned into simpler ones. Hence, *G* for the original geometry can be obtained by assembling *G* for all constituting partitions.

The following transformation proves to be useful for incompressible flow ($\nabla \cdot \mathbf{v}=0$):

$$G(\mathbf{r},\mathbf{r}') = e^{-\mathbf{v}\cdot(\mathbf{r}-\mathbf{r}')/2\alpha}\, g(\mathbf{r},\mathbf{r}') \quad \mathbf{r},\mathbf{r}'\in\Omega \quad (4.5)$$

where $\alpha = \lambda / \rho c_p$. Substituting in (4.2) yields:

$$\lambda\left[\nabla^2 g(\mathbf{r},\mathbf{r}') - \mathbf{v}\cdot\mathbf{v}\, g/4\alpha^2\right] = -\delta(\mathbf{r}-\mathbf{r}') \quad \mathbf{r},\mathbf{r}'\in\Omega \quad (4.6)$$

In case the velocity field was uniform, than (4.6) has a simple straightforward solution. For a complicated velocity field, *g* (or *G*) has to be obtained numerically.

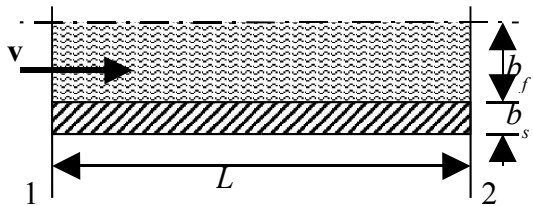

Figure 2. Problem geometry

## IV. A SIMPLE PROBLEM

The simple 2D channel will be considered here with symmetry boundary conditions at the channel mid plane, and uniform velocity field. Channel walls are of uniform thickness and physical properties. Both CTMs (for the conductive and the convective parts) are obtained directly from the available analytical Greens functions. These models are currently tested both as stand alone, as well as combined together to model conjugate heat transfer in order to prove the ability of the flexible profile approach to deal with such cases.

## V. CONCLUSION

A general formulation has been made of compact thermal models for convective domains that is fully compatible with that of conductive domains. The traditional heat transfer coefficient is now only a term in an influence matrix that correctly models heat transfer for different kinds of boundary conditions.

For the case of conjugate heat transfer in particular, we obtain the following advantages:

A. The problem becomes totally decoupled. We get a single model for the intrinsic behavior of the convective domain, as well as another single model for the conductive domain. Each of them depends only on its own domain characteristics and is *independent* of the other domain. They can be joined together to get the compound model that takes care of all interactions by simple algebraic operations.

B. The case of multiple heat sources is adequately treated due to the replacement of a single heat transfer coefficient by the new form of the CTM, which is a matrix of influence coefficients. The latter aspect can be viewed as a generalization of the concept of "adiabatic" heat transfer coefficient.

C. Surface temperature gradients are an outcome of the analysis, which will enable the estimation of thermal stresses.

D. Higher precision, as compared to using the classical heat transfer coefficient, can be obtained. This precision can be made as close as needed to that of the detailed FEM model, by increasing the number of elements in the basis. The key point is that convergence is very fast, hence few members are usually enough to get results of adequate precision.